\documentclass[runningheads,a4paper]{llncs}

\usepackage{amssymb,amsmath} 
\usepackage{graphicx}
\usepackage{cite}

\usepackage{url}
\urldef{\mailsa}\path|{m.baldi, m.bianchi, f.chiaraluce}@univpm.it|
\urldef{\mailsb}\path|rosenthal@math.uzh.ch|
\urldef{\mailsc}\path|davide.schipani@ntu.ac.uk|
\newcommand{\keywords}[1]{\par\addvspace\baselineskip
\noindent\keywordname\enspace\ignorespaces#1}

\begin{document}

\mainmatter  

\title{Using LDGM Codes and Sparse Syndromes \\ to Achieve Digital Signatures
\let\thefootnote\relax\footnotetext{This work was supported in part by the MIUR project ``ESCAPADE''
(Grant RBFR105NLC) under the ``FIRB -– Futuro in Ricerca 2010'' funding program, and in part by the Swiss National Science Foundation
under grant No. 132256.}
}

\titlerunning{Using LDGM Codes and Sparse Syndromes to Achieve Digital Signatures}

%
%
\author{Marco Baldi\inst{1} \and Marco Bianchi\inst{1} \and Franco Chiaraluce\inst{1} \and Joachim Rosenthal\inst{2} \and Davide Schipani\inst{3}}
\authorrunning{M. Baldi \and M. Bianchi \and F. Chiaraluce \and J. Rosenthal \and D. Schipani}

\institute{Universit\`a Politecnica delle Marche, Ancona, Italy\\ \mailsa\\
\and University of Zurich, Zurich, Switzerland\\ \mailsb \\
\and Nottingham Trent University, Nottingham, UK\\ \mailsc}

%
%

\toctitle{Lecture Notes in Computer Science}
\tocauthor{Authors' Instructions}
\maketitle

\begin{abstract}
In this paper, we address the problem of achieving efficient code-based digital signatures
with small public keys.
The solution we propose exploits sparse syndromes and randomly designed low-density
generator matrix codes.
Based on our evaluations, the proposed scheme is able to outperform existing solutions,
permitting to achieve considerable security levels with very small public keys.
\end{abstract}

\section{Introduction}

The problem of replacing current cryptographic primitives which will be subject to quantum computer attacks with alternative
post-quantum solutions is fostering research on code-based systems, which are among the most promising options for this replacement.

Concerning asymmetric cryptography, the McEliece cryptosystem \cite{McEliece1978} and its recent improvements \cite{Bernstein2008}
already represent efficient solutions to replace quantum vulnerable systems, like RSA.
The main drawback of the McEliece cryptosystem compared to RSA is the large size of its public keys.
However, great steps have been done towards the reduction of the McEliece public key size.
A possible solution consists in replacing the Goppa codes used in the original system with other families of codes.
Among these, low-density parity-check (LDPC) codes have been considered since several years \cite{Monico2000, Baldi2007ICC, Baldi2007ISIT, Baldi2008},
and most recent proposals based on them have been able to achieve significant reductions in the key size \cite{Baldi2012, Misoczki2012, Baldi2013}.

For what concerns digital signatures, the widespread DSA and RSA signature schemes will be endangered by quantum computers as well,
and only a few replacements are available up to now, like hash-based signatures.
Code-based digital signature schemes represent another post-quantum alternative to DSA and RSA signature schemes,
but the development of efficient code-based solutions is still challenging.

The two main proposals of code-based signature schemes currently available are the Courtois-Finiasz-Sendrier
(CFS) scheme \cite{Courtois2001} and the Kabatianskii-Krouk-Smeets (KKS) scheme \cite{Kabatianskii1997}.
An up-to-date discussion about these two systems can be found in \cite{Finiasz2011} and \cite{Otmani2011},
respectively.

The KKS scheme uses two codes with different sizes to create the trapdoor, one selecting the subset support of the other.
An important weakness of this system was recently pointed out in \cite{Otmani2011},
even though the KKS scheme can still be considered secure for some choices of its parameters.

The CFS signature scheme instead uses a hash-and-sign paradigm based on the fact that only the authorized
signer can exploit the error correction capability of a secret code. The main components of the CFS 
scheme are a private $t$-error correcting code $C$ and a public hash algorithm $\mathcal{H}$.
The private code is described through its parity-check matrix $H$, while $H' = S \cdot H$ is made public,
where $S$ is a private random matrix.
There must be a public function $\mathcal{F}$ able to transform (in a reasonable time) any hash value 
computed through $\mathcal{H}$ into a correctable syndrome for the code $C$.
Then, syndrome decoding through $C$ is performed by the signer, and the resulting error vector $e$
is the digital signature, together with the parameters to be used in the function $\mathcal{F}$
for achieving the target.
Verification is easily obtained by computing $H' \cdot e$ and comparing the result with the output of $\mathcal{F}$.

The main drawback of the CFS scheme concerns the function $\mathcal{F}$.
In fact, it is very hard to find a function that quickly transforms an arbitrary hash vector
into a correctable syndrome.
In the original CFS scheme, two ways are proposed to solve this problem \cite{Finiasz2011}:
$i$) appending a counter to the message, or $ii$) performing complete decoding.
Both these methods require a very special choice of the code parameters to be
able to find decodable syndromes within a reasonable time.
For this purpose, codes with very high rate and very small error correction capability
are commonly used, and this has exposed the cryptosystem to attacks based on the generalized 
birthday algorithm \cite{Finiasz2009}, in addition to common attacks against code-based cryptosystems.
This flaw is mainly due to the need to ensure that many vectors
produced by the hash function $\mathcal{H}$ are correctable.
In addition, in such a setting, the decoding complexity can be rather high,
especially in the versions exploiting complete decoding.

In this paper, we propose a new solution based on a modification of the CFS scheme.
The first variation is to consider only a subset of the possible syndromes,
selecting the ones having a certain density (of not null elements).
In addition, we replace traditional Goppa codes with low-density generator-matrix (LDGM)
codes, which allow for a random based design and a considerable reduction in the
public key size.
As it will be shown in the following, this allows to relax many constraints on the code parameters,
and therefore to use more practical codes which also make classical attacks against
the CFS scheme inapplicable.
In addition, syndrome decoding through the private code is reduced to a
straightforward procedure, with an extremely low complexity.
The rationale of the proposed system is in the following observations:
\begin{itemize}
\item Given a private parity-check matrix in systematic form (with an identity block in the rightmost part), the signer can obtain an error vector associated
to a given syndrome by simply transposing the syndrome and prepending it with an all zero vector.
By obtaining the public parity-check matrix from the private one through a left and right multiplication by two
dense secret matrices, the systematic form is lost and the same procedure cannot be exploited by an attacker.
Moreover, the two parity-check matrices no longer describe the same code.
\item The private error vector obtained by the signer can be disguised by adding to it a randomly selected codeword 
of the secret code.
\item If both the private error vector and the random codeword are of moderately low weight, and the same holds for 
their transposition into the public code, they are difficult to discover by an attacker.
\item If the private code is an LDGM code, it is very easy for the signer to randomly select a low weight codeword,
since it is obtained as the sum of a small number of rows of its generator matrix, chosen at random.
Although the private code is an LDGM code, its parity-check matrix in systematic form can be dense.
\end{itemize}

In the following sections we show how these observations are translated into practice in the proposed system.
The organization of the paper is as follows.
In Section \ref{sec:LDGMcodes}, we describe the LDGM codes we use in the system and their characteristics.
In Section \ref{sec:System}, we define the main steps of the system, that are: key generation, signing procedure and verification procedure.
In Section \ref{sec:Vulnerabilities}, we provide a preliminary assessment of the possible vulnerabilities affecting the system.
In Section \ref{sec:Examples}, we give some possible choices of the system parameters and, finally,
Section \ref{sec:Conclusion} concludes the paper.

\section{Low-density generator-matrix codes \label{sec:LDGMcodes}}

LDGM codes have been used since a long time for transmission applications \cite{Cheng1996},
and are recognized to achieve very good error correcting performance when used in concatenated
schemes \cite{Garcia-Frias2003}, \cite{Gonzalez-Lopez2007}.

A simple way to obtain an LDGM code with length $n$, dimension $k$ and redundancy $r=n-k$ is to define its generator matrix in the form
\begin{equation}
G = [I_k | D],
\label{eq:GLDGM}
\end{equation}
where $I_k$ is a $k \times k$ identity matrix, and $D$ is a sparse $k \times r$ matrix.
We suppose that the rows of $G$ have Hamming weight $w_g \ll n$.
An LDGM code can also be defined with $G$ in a more general form than \eqref{eq:GLDGM},
that is, by randomly selecting $k$ linearly independent vectors with length $n$ and 
Hamming weight $w_g \ll n$, and using them as the rows of $G$.
This approach requires to check the linear independence of the rows of $G$, but it
increases the degrees of freedom for random-based designs.
Hence, we consider this more general solution for the design of the private code in
the proposed system.

Due to their sparse nature, it is very likely that, by summing two or more rows of
the generator matrix of an LDGM code, a vector with Hamming weight $\ge w_g$ is obtained.
In this case, the LDGM code has minimum distance $w_g$. 
This is even more likely if the rows of $G$ are chosen in such a way as to be quasi-orthogonal,
that is, with a minimum number of overlapping ones.
However, in the scheme we propose, we do not actually need that the secret code has minimum
distance $w_g$. Hence, $G$ can be designed completely at random, without any constraint on the number 
of overlapping ones between each pair of its rows.

The code defined through $G$ as in \eqref{eq:GLDGM} is systematic and admits a sparse parity check matrix $H$ in the form
\begin{equation}
H = [D^T | I_r],
\end{equation}
where $^T$ denotes transposition and $I_r$ is an $r \times r$ identity matrix.
Hence, such a code is an LDPC code as well.
On the contrary, if $G$ is designed completely at random, without imposing the form \eqref{eq:GLDGM},
it is not systematic and the LDGM code is generally not an LDPC code.
This is the case for the private LDGM code which is used in the proposed system.

A particularly interesting class of LDGM codes is that of quasi-cyclic (QC) LDGM
codes \cite{Baldi2011IT}.
In fact, the QC property allows to reduce the memory needed to store the code
characteristic matrices, which is an important feature in cryptographic applications
where such matrices are used as private and public keys.

A general form for the generator matrix of a QC-LDGM code is as follows:
\begin{equation}
G_{QC} = \left[
\begin{array}{ccccc}
C_{0,0} & C_{0,1} & C_{0,2} & \ldots & C_{0,n_0-1} \\
C_{1,0} & C_{1,1} & C_{1,2} & \ldots & C_{1,n_0-1} \\
C_{2,0} & C_{2,1} & C_{2,2} & \ldots & C_{2,n_0-1} \\
\vdots & \vdots & \vdots & \ddots & \vdots \\
C_{k_0-1,0} & C_{k_0-1,1} & C_{k_0-1,2} & \ldots & C_{k_0-1,n_0-1} \\
\end{array}
\right],
\label{eq:GQCLDGM}
\end{equation}
where $C_{i,j}$ represents a sparse circulant matrix or a null matrix
with size $p \times p$.
Hence, in this case the code length, dimension and redundancy are
$n=n_0p$, $k=(n_0-r_0)p=k_0p$ and $r=r_0p$, respectively.
Since a circulant matrix is defined by one of its rows (conventionally the first),
storing a binary matrix $G_{QC}$ as in \eqref{eq:GQCLDGM} requires
$k_0 n_0 p = kn/p$ bits, and the corresponding parity-check matrix $H_{QC}$
requires $r_0 n_0 p = rn/p$ bits to be stored.
The proposed system uses a parity-check matrix as the public key; so, when
adopting QC-LDGM codes, its public key size is $rn/p$ bits.

An important feature of LDGM codes which will be exploited in the proposed
scheme is that it is easy to obtain a random codeword $c$ belonging to an
LDGM code and having weight approximately equal to a fixed, small value $w_c$.
Let us suppose that $w_c$ is an integer multiple of $w_g$.
Since the rows of $G$ are sparse, it is highly
probable that, by summing a small number of rows, the Hamming weight of the
resulting vector is about the sum of their Hamming weights.
Hence, by summing $\frac{w_c}{w_g}$ rows of $G$, chosen at random, we get
a random codeword with Hamming weight about $w_c$.
Actually, due to some overlapping ones, the resulting weight could result
smaller than $w_c$. In this case, some other row can be added, or some row
replaced, or another combination of rows can be tested, in order to approach $w_c$.
Moreover, as we will see in the following, using a random 
codeword with Hamming weight slightly smaller than $w_c$ is not a problem in the proposed system.
Based on the above considerations, the number of codewords with weight close to $w_c$
which can be easily selected at random from an LDGM code having $G$ with rows of weight $w_g$,
with $w_g | w_c$, can be roughly estimated as
\begin{equation}
A_{w_c} \approx {k \choose \frac{w_c}{w_g}}.
\end{equation}

\section{System description}
\label{sec:System}

In this section we describe the main steps of the proposed digital signature system.

\subsection{Key generation}

The first part of the private key for the proposed system is formed by the $r \times n$ parity-check matrix $H$
of an LDGM code $C(n,k)$, having length $n$ and dimension $k$ ($r = n-k$).
The matrix $H$ is in systematic form, with an identity block in the rightmost part.
The private key also includes two other non-singular matrices: an $r \times r$ transformation matrix $Q$ and an $n \times n$
scrambling matrix $S$ (both defined below).
The public key is then obtained as $H' = Q^{-1} \cdot H \cdot S^{-1}$.

The matrix $S$ is a sparse non-singular matrix, with average row and column weight $m_S \ll n$.
The matrix $Q$, instead, is a \textit{weight controlling} transformation matrix as defined in \cite{Baldi2011}.
For this kind of matrices, when $s$ is a suitably chosen sparse vector, the vector $s' = Q \cdot s$ has a small Hamming weight,
which is only a few times greater than that of $s$.
As shown in \cite{Baldi2011}, a matrix $Q$ with such a feature can be obtained as the sum of an $r \times r$
low-rank dense matrix $R$ and a sparse matrix $T$, chosen in such a way that $Q = R+T$ is non singular.
In order to design $R$, we start from two $z \times r$ matrices, $a$ and $b$, with $z<r$ properly chosen (see below),
and define $R$ as:
\begin{equation}
R = a^T \cdot b.
\label{eq:R}
\end{equation}
This way, $R$ has rank $\le z$.
The matrix $T$ is then chosen as a sparse matrix with row and column weight $m_T$, such that
$Q = R+T$ is full rank.

It can be easily verified that, if the vector $s$ is selected in such a way that $b \cdot s = 0_{z \times 1}$,
where $0_{z \times 1}$ is the $z \times 1$ all-zero vector, then $R \cdot s = 0_{r \times 1}$ and $s' = Q \cdot s = T \cdot s$.
Hence, the Hamming weight of $s'$ is, at most, equal to $m_T$ times that of $s$, and $Q$ actually has the
weight controlling feature we desire.

As we will see in Section \ref{subsec:KeyRecAtt}, although it is relatively simple for an attacker to obtain the space 
defined by the matrix $b$, and its dual space, this does not help to mount a key recovery attack.
Hence, the matrix $b$, which is only a small part of $Q$, can even be made public.

When a QC code is used as the private code, $H$ is formed by $r_0 \times n_0$ circulant matrices of size $p \times p$,
and it is desirable to preserve this QC structure also for $H'$, in such a way as to exploit its benefits in terms of key size.
For this purpose, both $Q$ and $S$ must be formed by circulant blocks with the same size as those forming $H$.
Concerning the matrix $S$, it is obtained in QC form ($S_{QC}$) by simply choosing at random
a block of $n_0 \times n_0$ sparse or null circulant matrices such that the overall row and column weight is $m_S$.

Concerning the matrix $Q$, instead, a solution to obtain it in QC form is to define $R$ as follows:
\begin{equation}
R_{QC} = \left( a_{r_0}^T \cdot b_{r_0} \right) \otimes \mathbf{1}_{p \times p},
\label{eq:RQC}
\end{equation}
where $a_{r_0}$ and $b_{r_0}$ are two $z \times r_0$ binary matrices, $\mathbf{1}_{p \times p}$ is the all-one $p \times p$
matrix and $\otimes$ denotes the Kronecker product.
Then, $T_{QC}$ is chosen in the form of $n_0 \times n_0$ sparse circulant blocks with overall row and column weight $m_T$
and $Q_{QC}$ is obtained as $R_{QC} + T_{QC}$.
This way, if $H$ is in QC form, $H' = Q_{QC}^{-1} \cdot H \cdot S_{QC}^{-1}$ is in QC form as well.
In the QC case, the condition we impose on $s$, that is, $b \cdot s = 0_{z \times 1}$ becomes
$\left( b_{r_0} \otimes \mathbf{1}_{1 \times p} \right) \cdot s = 0_{z \times 1}$.

Such a condition, both in the generic and in the QC case, is equivalent to a set of $z$ parity-check constraints
for a code with length $r$ and redundancy $z$.
Hence, if we fix $b$ such that this code has minimum distance $d$, then a vector $s$ with weight $w < d$ cannot
satisfy such condition, and $Q$ loses its weight controlling feature on $s$.
This is useful to reinforce the system against some vulnerabilities, and justifies the form used for the matrix $Q$.

Apart from the private and public key pair, the system needs two functions which are made public as well:
a hash function $\mathcal{H}$ and a function $\mathcal{F}_\Theta$ that converts the output vector of $\mathcal{H}$ into
a sparse $r$-bit vector $s$ of weight $w \ll r$.
The output of $\mathcal{F}_\Theta$ depends on the parameter $\Theta$, which is associated to the message to be signed
and made public by the signer.
An example of implementation of $\mathcal{F}_\Theta$ is provided in the next section.

\subsection{Signature generation}

In order to get a unique digital signature from some document $M$, the signer
computes the digest $h = \mathcal{H}(M)$ and then finds $\Theta_M$
such that $s = \mathcal{F}_{\Theta_M}(h)$ verifies $b \cdot s = 0_{z \times 1}$.
Since $s$ has weight $w$, $s' = Q \cdot s$ has weight $\le m_T w$.
Concerning the implementation of the function $\mathcal{F}_{\Theta}(h)$, an example 
is as follows.
Given a message digest $h = \mathcal{H}(M)$ of length $x$ bits, similarly to what is done in the CFS scheme, it is appended with the $y$-bit 
value $l$ of a counter, thus obtaining $[h | l]$.
The value of $[h | l]$ is then mapped uniquely into one of the ${r \choose w}$ $r$-bit vectors of weight $w$, hence it must
be ${r \choose w} \ge 2^{x+y}$.
The counter is initially set to zero by the signer, and then progressively increased.
This way, a different $r$-bit vector is obtained each time, until one orthogonal to $b$ is found, for $l = \bar{l}$.
This step requires the signer to test $2^{z}$ different values of the counter, on average.
With this implementation of $\mathcal{F}_{\Theta}(h)$, we have $\Theta_M = \bar{l}$, and different signatures correspond 
to different vectors $s$, unless a hash collision occurs.

After having obtained $s$, the signer has to find a vector $e$ of weight $\le m_T w$ which corresponds to the private syndrome $s' = Q \cdot s$ through $C$.
Since $H$ is in systematic form, it can be written as $H = [X|I_r]$, where $X$ 
is an $r \times k$ matrix and $I_r$ is the $r \times r$ identity matrix.
Hence, the private syndrome $s'$ can be obtained from the error vector $e = [0_{1 \times k}|s'^T]$.
So, in this case, finding $e$ simply translates into a vector transposition and some zero padding.

The signer finally selects a random codeword $c \in C$ with small Hamming weight ($w_c$),
and computes the public signature of $M$ as $e' = (e+c) \cdot S^T$.
If the choice of the codeword $c$ is completely random and independent of the document to be signed, the signature obtained for 
a given document changes each time it is signed, and the system becomes vulnerable to attacks exploiting many signatures of the same document.
This can be simply avoided by choosing the codeword $c$ as a deterministic function of the document $M$ and, hence, of the public syndrome $s$.
For example, $s$ or, equivalently, $[h|\bar{l}]$ can be used as the initial state of the pseudo-random integer generator through which 
the signer extracts the indexes of the rows of $G$ that are summed to obtain $c$.
This way, the same codeword is always obtained for the same public syndrome.

To explain the role of the codeword $c$, let us suppose for a moment that the system does not include any random codeword,
that is equivalent to fix $c = 0_{1 \times n}, \forall M$.
In this case, we could write $e' = W(s)$, where $W$ is a linear bijective map from the set
of public syndromes to the set of valid signatures. This can be easily verified, since
it is simple to check that $W(s_1 + s_2) = W(s_1) + W(s_2)$.
So, an attacker who wants to forge a signature for the public syndrome $s_x$
could simply express $s_x$ as a linear combination of previously intercepted
public syndromes, $s_x = s_{i_1} + s_{i_2} + \ldots s_{i_N}$, and forge a valid 
signature by linearly combining their corresponding signatures:
$e'_x = e'_{i_1} + e'_{i_2} + \ldots e'_{i_N}$.

As mentioned, to prevent this risk, the signer adds a random codeword $c$, with weight $w_c \ll n$,
to the error vector $e$, before multiplication by $S^T$.
This way, the map $W$ becomes an affine map which depends on the random codeword $c$,
and it no longer has the set of valid signatures as its image.
In fact, we can denote this new map as $W_c(s)$, such that $e_1' = W_{c_1}(s_1)$
and $e_2' = W_{c_2}(s_2)$, where $c_1$ and $c_2$ are two randomly selected codewords
of the private code with weight $w_c$.
If we linearly combine the signatures, we obtain $e_f' = e_1' + e_2' = W_{c_1}(s_1) + W_{c_2}(s_2) = W_{c_1 + c_2}(s_1 + s_2)$.
The vector $c_1 + c_2$ is still a valid codeword of the secret code, but it has weight $> w_c$
with a very high probability.

\subsection{Signature verification}

After receiving the message $M$, its signature $e'$ and the associated parameter $\Theta_M$,
the verifier first checks that the weight of $e'$ is $\le (m_Tw+w_c)m_S$.
If this condition is not satisfied, the signature is discarded.
Then the verifier computes $\widehat{s} = \mathcal{F}_{\Theta_M} \left( \mathcal{H}(M) \right)$
and checks that $\widehat{s}$ has weight $w$, otherwise the signature is discarded.
If the previous checks have been positive, the verifier then computes $H' \cdot e'^T = Q^{-1} \cdot H \cdot S^{-1} \cdot S \cdot (e^T + c^T) = Q^{-1} \cdot H \cdot (e^T + c^T) = Q^{-1} \cdot H \cdot e^T = Q^{-1} \cdot s' = s$.
If $ s = \widehat{s}$, the signature is accepted; otherwise, it is discarded.

\subsection{Number of different signatures}

An important parameter for any digital signature scheme is the total number
of different signatures.
In our case, a different signature corresponds to a different $r$-bit vector
$s$, having weight $w$.
Only vectors $s$ satisfying the $z$ constraints imposed by $b$ are
acceptable, so the maximum number of different signatures is:
\begin{equation}
N_s \approx \frac{{r \choose w}}{2^z}.
\end{equation}

\section{Possible vulnerabilities}
\label{sec:Vulnerabilities}

For a security assessment of the proposed system, it would be desirable to find possible
security reductions to some well known hard problems, and then to evaluate the complexity
of practical attacks aimed at solving such problems.
This activity is still at the beginning, and work is in progress in this direction.
Hence, in this paper we only provide a sketch of some possible vulnerabilities we
have already devised, which permit to obtain a first rough estimate of the security
level of the system.
Completing the security assessment will allow to improve the security level estimation,
and possibly to find more effective choices of the system parameters.

From the definition of the proposed system, it follows that the published signature $e'$
associated to a document $M$ is always a sparse vector, with Hamming weight $\le (m_Tw+w_c)m_S$.
Since $e'$ is an error vector corresponding to the public syndrome $s$ through the public code
parity-check matrix $H'$, having a low Hamming weight ensures that $e'$ is difficult to find,
starting from $s$ and $H'$.
This is achieved by using the weight controlling matrix $Q$ and the sparse matrix $S$.
If this was not the case, and $e'$ was a dense vector, it would be easy to forge signatures,
since a dense vector corresponding to $s$ through $H'$ is easy to find.

Based on these considerations, one could think that choosing both $Q$ and $S$ as sparse as possible
would be a good solution.
Let us suppose that they are two permutation matrices, $P_1$ and $P_2$.
In this case, the public matrix would be $H' = P_1^T \cdot H \cdot P_2^T$, and
both $s'$ and $e'$ would be sparse, thus avoiding easy forgeries.
Actually, a first reason for avoiding to use permutation matrices is that, when masked 
only through permutations, the security of $H$ decreases.
In fact, using a doubly permuted version of $H$ may still allow to perform 
decoding through the public code.
However, neglecting for a moment this fact, we find that, in this case, $e'$ would have weight $\le w + w_c$. 
If $e$ and $c$ have disjoint supports, which is very likely true, since
we deal with sparse vectors, $w$ non-zero bits in $e'$ would correspond to a reordered
version of the non-zero bits in $s$.
So, apart from the effect of the random codeword, we would simply have
a disposition of the non-zero bits in $s$ within $e'$, according to a fixed pattern.
This pattern could be discovered by an attacker who observes a
sufficiently large number of signatures, so that the effect
of the random codeword could be eliminated.
In fact, by computing the intersection of the supports of many vectors
$s$ and their corresponding vectors $e'$, the support of $e'$ could be decomposed
and the reordering of each bit disclosed.

Based on these considerations, we can conclude that the density of $e'$ must be carefully chosen between two opposite needs:
\begin{itemize}
\item being sufficiently low to avoid forgeries;
\item being sufficiently high to avoid support decompositions.
\end{itemize}

\subsection{Forgery attacks}

In order to forge signatures, an attacker could search for an $n \times r$ 
right-inverse matrix $H'_r$ of $H'$. 
Then, he could compute $f = (H'_r \cdot s)^T$, which is a forged signature.
It is easy to find a right-inverse matrix able to forge dense signatures.
In fact, provided that $H' \cdot H'^T$ is invertible, 
$H'_r = H'^T \cdot \left(H' \cdot H'^T\right)^{-1}$ is a right-inverse matrix of $H'$.
The matrix $H'$ is dense and the same occurs, in general, for $\left(H' \cdot H'^T\right)^{-1}$;
so, $H'_r$ is dense as well. It follows that, when multiplied by $s$, $H'_r$ produces a dense
vector, thus allowing to forge dense signatures.
By using sparse signatures, with weight $\le (m_Tw + w_c)m_S$, the proposed system is
robust against this kind of forged signatures.

However, the right-inverse matrix is not unique.
So, the attacker could search for an alternative, possibly sparse, right-inverse matrix.
In fact, given an $n \times n$ matrix $Z$ such that $H' \cdot Z \cdot H'^T$ is invertible,
$H''_r = Z \cdot H'^T \cdot \left(H' \cdot Z \cdot H'^T\right)^{-1}$
is another valid right-inverse matrix of $H'$.
We notice that $H''_r \ne Z \cdot H'_r$.
When $H'$ contains an invertible $r\times r$ square block, there is also another simple way to find 
a right-inverse.
It is obtained by inverting such 
block, putting its inverse at the same position (in a transposed matrix)
in which it is found within $H'$, and padding the remaining rows with zeros.

In any case, there is no simple way to find a right-inverse matrix that is also sparse, 
which is the aim of an attacker.
Actually, for the matrix sizes considered here, the number of possible 
choices of $Z$ is always huge.
Moreover, there is no guarantee that any of them produces a sparse right-inverse.
Searching for an $r \times r$ invertible block within $H'$ and inverting it would also
produce unsatisfactory results, since the overall density of $H'^{-1}$ is reduced,
but the inverse of the square block is still dense.
So, the attacker would be able to forge signatures with a number of symbols $1$
on the order of $r/2$, that is still too large for the system considered here.
In fact, in the system examples we provide in Section \ref{sec:Examples}, we
always consider public signatures with weight on the order of $r/3$ or less.

A further chance is to exploit Stern's algorithm \cite{Stern1989} (or other approaches for
searching low weight codewords) to find a sparse representation of the
column space of $H'_r$. If this succeeds, it would result in a sparse matrix 
$H_S = H'_r \cdot B$, for some $r \times r$ transformation 
matrix $B$.
However, in this case, $H_S$ would not be a right-inverse of $H'$.

For these reasons, approaches based on right-inverses seem to be infeasible for
an attacker.
An alternative attack strategy could be based on decoding.
In fact, an attacker could try syndrome decoding of $s$ through $H'$, hoping
to find a sparse vector $f$.
He would have the advantage of searching for one out of many possible vectors,
since he is not looking for a correctable error vector.
Several algorithms could be exploited for solving such problem \cite{Chabaud96, Peters2010, May2011, Bernstein2011, Becker2012}.
These algorithms are commonly used to search for low weight
vectors with a null syndrome, but, with a small modification, they can also be used
to find vectors corresponding to a given (non-zero) syndrome.
In addition, their complexity decreases when an attacker has access to a high number of decoding instances,
and wishes to solve only one of them \cite{Sendrier2011}, which is the case for the proposed system.
We will discuss the complexity issue in Section \ref{sec:Examples}.

\subsection{Support decomposition attacks}

Concerning support decomposition, let us suppose that $e$ and $c$ have disjoint supports.
In this case, the overall effect of the proposed scheme on the public syndrome $s$
can be seen as the expansion of an $r \times 1$ vector $s$ of weight $w$ into
a subset of the support of the $1 \times n$ vector $e'$, having weight $\le m_T m_S w$,
in which each symbol $1$ in $s$ corresponds, at most, to $m = m_T m_S$ symbols $1$ in $e'$.

An attacker could try to find the $w$ sets of $m$ (or less) symbols $1$ within 
the support of $e'$ in order to compute valid signatures.
In this case, he will work as if the random codeword was absent, that is, $c = 0_{1 \times n}$.
Thus, even after succeeding, he would be able to forge signatures that are sparser
than the authentic ones.
In any case, this seems a rather dangerous situation, so we should aim at designing
the system in such a way as to avoid its occurrence.

To reach his target, the attacker must collect a sufficiently large number $L$ of pairs $(s, e')$. 
Then, he can intersect the supports (that is, compute the bit-wise AND) of all the $s$ vectors.
This way, he obtains a vector $s_L$ that may have a small weight $w_L \ge 1$.
If this succeeds, the attacker analyzes the vectors $e'$, and selects the $m w_L$
set bit positions that appear more frequently.
If these bit positions are actually those corresponding to the $w_L$ bits set in $s_L$,
then the attacker has discovered the relationship between them.
An estimate of the probability of success of this attack can be obtained through
combinatorial arguments.

An even more efficient strategy could be to exploit information set decoding to
remove the effect of the random codeword.
In fact, an attacker knows that $e' = (e+c) \cdot S^T = e'' + c''$, with $c''$ such that $H'c'' = 0$.
Hence, $e''$ can be considered as an error vector with weight $\le m_T m_S w$ affecting
the codeword $c''$ of the public code.
So the attacker could consider a random subset of $k$ coordinates of the public
signature $e'$ and assume that no errors occurred on these coordinates.
In this case, he can easily recover $c''$ and, hence, remove the effect of the random codeword $c$.
The probability that there are no errors in the chosen $k$ coordinates is ${n - m_T m_S w \choose k}/{n \choose k}$,
and its inverse provides a rough estimate of the work factor of this attack.

\subsection{Key recovery attacks}
\label{subsec:KeyRecAtt}

An attacker could aim to mount a key recovery attack, that is, to obtain the private code.
A potential vulnerability in this sense comes from the use of LDGM codes.
As we have seen in Section \ref{sec:LDGMcodes}, LDGM codes offer the advantage of having a predictable (and sufficiently
high) number of codewords with a moderately low weight $w_c$, and of making
their random selection very easy for the signer.
On the other hand, when the private code is an LDGM code, the public code admits a 
generator matrix in the form $G_I' = G \cdot S^T$, which is still rather sparse.
So, the public code contains low weight codewords, coinciding with the rows of $G_I'$,
which have weight approximately equal to $w_g \cdot m_S$.
Since $G_I'$ has $k$ rows, and summing any two of them gives higher weight codewords
with a very high probability, we can consider that the multiplicity of these
words in the public code is $k$.
They could be searched by using again Stern's algorithm \cite{Stern1989} and
its improved versions \cite{Chabaud96, Peters2010, May2011, Bernstein2011, Becker2012}, in such a way as to recover $G_I'$.
After that, $G_I'$ could be separated into $G$ and $S^T$ by exploiting their sparsity.
In Section \ref{sec:Examples} we discuss how to estimate the work factor of this attack.

Another possible vulnerability comes from the fact that the matrix $b$ is public.
Even if $b$ was not public, an attacker could obtain the vector space generated by $b$, as well as its dual space,
by observing $O(r)$ public syndromes $s$, since $b \cdot s = 0_{z \times 1}$.
Hence, we must suppose that an attacker knows an $r \times r$ matrix $V$ such that $R \cdot V = 0 \Rightarrow Q \cdot V = T \cdot V$.
The attacker also knows that $H' = Q^{-1} \cdot H \cdot S^{-1}$ and that the public code admits any non-singular generator matrix in the form
$G_X' = X \cdot G \cdot S^T$, which becomes $G_Q' = Q \cdot G \cdot S^T$ for $X = Q$.
Obviously, $G_I'$ is the sparsest among them, and it can be attacked by searching for low weight codewords in the public code,
as we have already observed.
Instead, knowing $V$ is useless to reduce the complexity of attacking either $H'$ or one of the possible $G_X'$,
hence it cannot be exploited by an attacker to perform a key recovery attack.

\subsection{Other attacks}

As for any other hash-and-sign scheme, classical collision birthday attacks represent a threat
for the proposed system.
Since the system admits up to $N_s$ different signatures, it is sufficient to collect $\approx \sqrt{N_s}$
different signatures to have a high probability of finding a collision \cite{Lim1995}.
Hence, the security level reached by the system cannot exceed $\sqrt{N_s}$.

However, $N_s$ can be made sufficiently high by increasing the value of $w$.
The definition of the proposed system allows to choose its parameters in such a way as to guarantee this fact,
as we will see in Section \ref{sec:Examples}.
This is possible since the choice of $w$ is not constrained by the row weight of the private generator matrix.
In fact, in the proposed scheme we do not actually need a private code of minimum distance greater than $2w$,
because we rely on a decoding procedure which uniquely associates to a syndrome
of a given weight an error vector with the same weight, though such an error vector is not necessarily unique.

Finally, it is advisable to consider the most dangerous attacks against the CFS scheme.
It was successfully attacked by exploiting syndrome decoding based on the generalized birthday algorithm \cite{Finiasz2009},
even if the proposed attacking algorithm was not the optimal generalization of the birthday algorithm \cite{Minder2012}.
If we do not take into account some further improvement due to the QC structure of the public key, these algorithms
provide a huge work factor for the proposed system parameters, since they try to solve the decoding problem for a random code.
Just to give an idea, we obtain a work factor of more than $2^{200}$ binary operations even for the smallest key sizes we consider.
However, there are some strategies that can be implemented to improve the efficiency of the attack 
on structured matrices, like those of dyadic codes \cite{Niebuhr2010}.
This improvement could be extended to QC codes as well, but the attack work factor, for the cryptosystems analyzed in \cite{Niebuhr2010},
is lowered by (at most) $2^{10}$ binary operations, starting from a maximum value of $2^{344}$.
Hence, it is very unlikely that this strategy can endanger the signature scheme we propose.

\section{System examples \label{sec:Examples}}

By using the preliminary security assessment reported in Section \ref{sec:Vulnerabilities},
we can find some possible choices of the system parameters aimed at reaching fixed
security levels. For this purpose, we have considered all the 
vulnerabilities described in Section \ref{sec:Vulnerabilities}, and we have
estimated the work factor needed to mount a successful attack exploiting each
of them.

We have used the implementation proposed in \cite{Peters2010} for estimating
the work factor of low weight codeword searches.
Actually, \cite{Peters2010} does not contain the most up-to-date and efficient implementation
of information set decoding.
In fact, some improvements have appeared in the literature concerning algorithms for decoding binary random codes (as \cite{May2011}, \cite{Becker2012}).
These papers, however, aim at finding algorithms which are asymptotically faster, by minimizing their asymptotic complexity exponents.
Instead, for computing the work factor of attacks based on decoding, we need actual operation counts, which are not reflected in these recent works.
Also ``ball collision decoding'' \cite{Bernstein2011} achieves significant improvements asymptotically, but they become negligible
for finite code lengths and not too high security levels.
For these reasons, we prefer to resort to \cite{Peters2010}, which provides a detailed analysis of the algorithm, together with a precise operation
count for given code parameters.
On the other hand, attacks against the proposed system which exploit decoding, i.e., trying to recover the rows of $G$ or to forge valid
signatures through decoding algorithms, are far away from providing the smallest work factors, and, hence, to determine the security level.
For the choices of the system parameters we suggest, the smallest work factor is always achieved by attacks aiming at decomposing the
signature support, which hence coincides with the security level.
For the instances proposed in this section, the work factor of attacks based on decoding is on the order of $2^{2 SL}$, where $SL$ is the claimed security level.
Hence, even considering some reduction in the work factor of decoding attacks would not change the security level of the considered instances of the system.
This situation does not change even if we consider the improvement coming from the ``decoding one out of many'' approach \cite{Sendrier2011}. 
In fact, as shown in \cite{Sendrier2011}, even if an attacker has access to an unlimited number of decoding instances, the attack complexity
is raised by a power slightly larger than $2/3$.

Concerning support decomposition attacks, a rough estimation of their complexity has instead
been obtained through simple combinatorial arguments, which are not reported here for the sake
of brevity.

A more detailed analysis of the attacks work factor is out of scope of
this paper, and will be addressed in future works, together with a more complete
security assessment. This will also permit to refine the choice of the system
parameters, in such a way as to find the best trade-off between the security level
and the key size.

Table \ref{tab:Examplesd2} provides three choices of the system parameters
which are aimed at achieving 80-bit, 120-bit and 160-bit security, respectively.
All these instances of the system use QC-LDGM codes with different values
of $p$, also reported in the table, and consider the case in which the matrix
$Q$ is such that $d= w_L = 2$.
Actually, achieving minimum distance equal to $2$ (or more) is very easy:
it is sufficient to choose $z > 1$ and to guarantee that the matrix $b$ does 
not contain all-zero columns.
For each instance of the system, the value of the key size $S_k$ is also shown,
expressed in kibibytes (1 KiB $= 1024 \cdot 8$ bits).

\begin{table*}[!t]
\renewcommand{\arraystretch}{1.3}
\caption{System parameters for some security levels (SL), with $d = 2$ and $w_L = 2$.}
\label{tab:Examplesd2}
\centering
\begin{tabular}{@{\hspace{0.8mm}}c@{\hspace{0.8mm}}|@{\hspace{0.8mm}}c@{\hspace{0.8mm}}|@{\hspace{0.8mm}}c@{\hspace{0.8mm}}|@{\hspace{0.8mm}}c@{\hspace{0.8mm}}|@{\hspace{0.8mm}}c@{\hspace{0.8mm}}|@{\hspace{0.8mm}}c@{\hspace{0.8mm}}|@{\hspace{0.8mm}}c@{\hspace{0.8mm}}|@{\hspace{0.8mm}}c@{\hspace{0.8mm}}|@{\hspace{0.8mm}}c@{\hspace{0.8mm}}|@{\hspace{0.8mm}}c@{\hspace{0.8mm}}|@{\hspace{0.8mm}}c@{\hspace{0.8mm}}|@{\hspace{0.8mm}}c@{\hspace{0.8mm}}|@{\hspace{0.8mm}}c@{\hspace{0.8mm}}}
SL (bits) & $n$ & $k$ & $p$ & $w$ & $w_g$ & $w_c$ & $z$ & $m_T$ & $m_S$ & $A_{w_c}$ & $N_s$ & $S_k$ (KiB) \\
\hline
\hline
80 & 9800  & 4900 & 50 & 18 & 20 & 160 & 2 & 1 & 9 & $2^{82.76}$ & $2^{166.10}$ & 117 \\
\hline
120 & 24960  & 10000 & 80 & 23 & 25 & 325 & 2 & 1 & 14 & $2^{140.19}$ & $2^{242.51}$ & 570 \\
\hline
160 & 46000 & 16000 & 100 & 29 & 31 & 465 & 2 & 1 & 20 & $2^{169.23}$ & $2^{326.49}$ & 1685 \\
\hline
\end{tabular}
\end{table*}

In the original version of the CFS system,
to achieve an attack time and memory complexity greater than $2^{80}$,
we need to use a Goppa code with length $n = 2^{21}$ and redundancy
$r = 21 \cdot 10 = 210$ \cite{Finiasz2009}. This gives a key size
on the order of $4.4 \cdot 10^8$ bits $=  52.5$ MiB.
By using the parallel CFS proposed in \cite{Finiasz2011}, the same
work factor can be reached by using keys with size ranging between
$1.05 \cdot 10^7$ and $1.7 \cdot 10^8$ bits, that is, between $1.25$ MiB and $20$ MiB.
As we notice from Table \ref{tab:Examplesd2}, the proposed system requires 
a public key of only $117$ KiB to achieve 80-bit security.
Hence, it is able to achieve a dramatic reduction in the
public key size compared to the CFS scheme, even when using a parallel 
implementation of the latter.


Another advantage of the proposed solution compared to the CFS scheme is that
it exploits a straightforward decoding procedure for the secret code.
On the other hand, $2^z$ attempts are needed, on average, to find an
$s$ vector such that $b \cdot s = 0_{z \times 1}$. However, such a check is very 
simple to perform, especially for very small values of $z$, like those considered here.

\section{Conclusion \label{sec:Conclusion}}

In this paper, we have addressed the problem of achieving efficient code-based
digital signatures with small public keys.

We have proposed a solution that, starting from the CFS schemes, exploits
LDGM codes and sparse syndromes to achieve good security levels with compact keys.
The proposed system also has the advantage of using a
straightforward decoding procedure, which reduces to a transposition
and a concatenation with an all-zero vector. This is considerably
faster than classical decoding algorithms used for common families 
of codes.

The proposed scheme allows to use a wide range of choices of the code parameters.
In particular, the low code rates we adopt avoid some problems of the classical CFS scheme,
due to the use of codes with high rate and small correction capability.

On the other hand, using sparse vectors may expose the system to new attacks.
We have provided a sketch of possible vulnerabilities affecting this system,
together with a preliminary evaluation of its security level.
Work is in progress to achieve more precise work factor estimates for the
most dangerous attacks.

\end{document}